\documentclass[twocolumn,trackchanges]{aastex631}
\usepackage{gensymb}
\usepackage{natbib}
\usepackage{float}
\usepackage{footnotebackref}
\usepackage{tablefootnote}
\usepackage{wrapfig}
\usepackage{listings}
\usepackage{placeins}
\usepackage{hyperref}
\usepackage[tight,footnotesize]{subfigure}

\shorttitle{Water Vapor Emission in Near-Infrared Enceladus Plume Spectra}
\shortauthors{Denny et al.}

\begin{document}

\title{Constraining Time Variations in Enceladus' Water-Vapor Plume With Near-Infrared Spectra from Cassini-VIMS}
\author{K.E. Denny}
\affiliation{Department of Physics, University of Idaho, Moscow, ID 83844, USA}

\author{M.M. Hedman}
\affiliation{Department of Physics, University of Idaho, Moscow, ID 83844, USA}

\author{D. Bockel{\'e}e-Morvan}
\affiliation{Observatoire de Paris, Universit\'e Paris-Diderot, Paris, 75010, France}

\author{G. Filacchione}
\affiliation{Istituto di Astrofisica e Planetologia Spaziali, Rome, 00133, Italy}

\author{F. Capaccioni}
\affiliation{Istituto di Astrofisica e Planetologia Spaziali, Rome, 00133, Italy}

\begin{abstract}

Water vapor produces a series of diagnostic emission lines in the near infrared between 2.60 and 2.75 microns. The Visual and Infrared Mapping Spectrometer (VIMS) onboard the Cassini spacecraft detected this emission signal from Enceladus' plume, and so VIMS observations provide information about the variability of the plume's water vapor content. Using a data set of 249 spectral cubes with relatively high signal-to-noise ratios, we confirmed the strength of this water-vapor emission feature corresponds to a line-of-sight column density of order $10^{20}$ $molecules/m^{2}$, which is consistent with previous measurements from Cassini's Ultraviolet Imaging Spectrograph (UVIS). Comparing observations made at different times indicates that the water-vapor flux is unlikely to vary systematically with Enceladus' orbital phase, unlike the particle flux, which does vary with orbital phase. However, variations in the column density on longer and shorter timescales cannot be ruled out and merit further investigation. 

\end{abstract}

 
\section{Introduction} \label{sec:intro}

Saturn's moon Enceladus is an exceptionally interesting object because it is geologically active, producing a plume of predominantly water-ice grains. Multiple instruments aboard Cassini observed both the particulate and vapor components of the plume material emerging from four fissures (tiger stripes) located near the south pole of the moon \citep[see][for recent reviews]{schenk_18}. Observations of this material have aided in the understanding of the geological processes occurring on Enceladus, however, there are still many questions about how the plume's activity varies over time.

Variations in Enceladus' activity on various timescales can provide insights into what is happening beneath the moon's surface. Fluctuations in the plume particle flux have been extensively studied using Cassini's Visual and Infrared Mapping Spectrometer (VIMS) \citep{2013Natur.500..182H, Sharma_23} and Imaging Science Subsystem (ISS) data \citep{2014AJ....148...46N,2020Icar..34413345I}. These studies show that the overall brightness and particle output of the plume is correlated with Enceladus' position along its eccentric orbit. Specifically, the plume's brightness peaks when Enceladus is furthest from Saturn in its orbit \citep{2013Natur.500..182H,2020Icar..34413345I}. These trends indicate that the particle flux is correlated with the tidal stresses experienced by Enceladus' South Polar Terrain \citep{hurford_09,2014AJ....148...46N}.

The plume's water-vapor variations have been significantly more difficult to document because the water vapor is most clearly observed in relatively short and infrequent observations like occultations and close flybys. This has led to a debate about whether the vapor plume varies on orbital timescales correlated with the particulate component or if it is relatively constant. Occultation data obtained by Cassini's Ultraviolet Imaging Spectrograph (UVIS) indicates that there is a constant vapor flux throughout Enceladus' orbit \citep{2020Icar..34413461H}. However, data collected by Cassini's Ion and Neutral Mass Spectrometer (INMS) during close flybys provides evidence of variations in the plume's gas jets \citep{Saur_2008,Smith_2010,2017AsBio..17..926T}.

\begin{figure*}[htb]
\includegraphics[width=\textwidth]{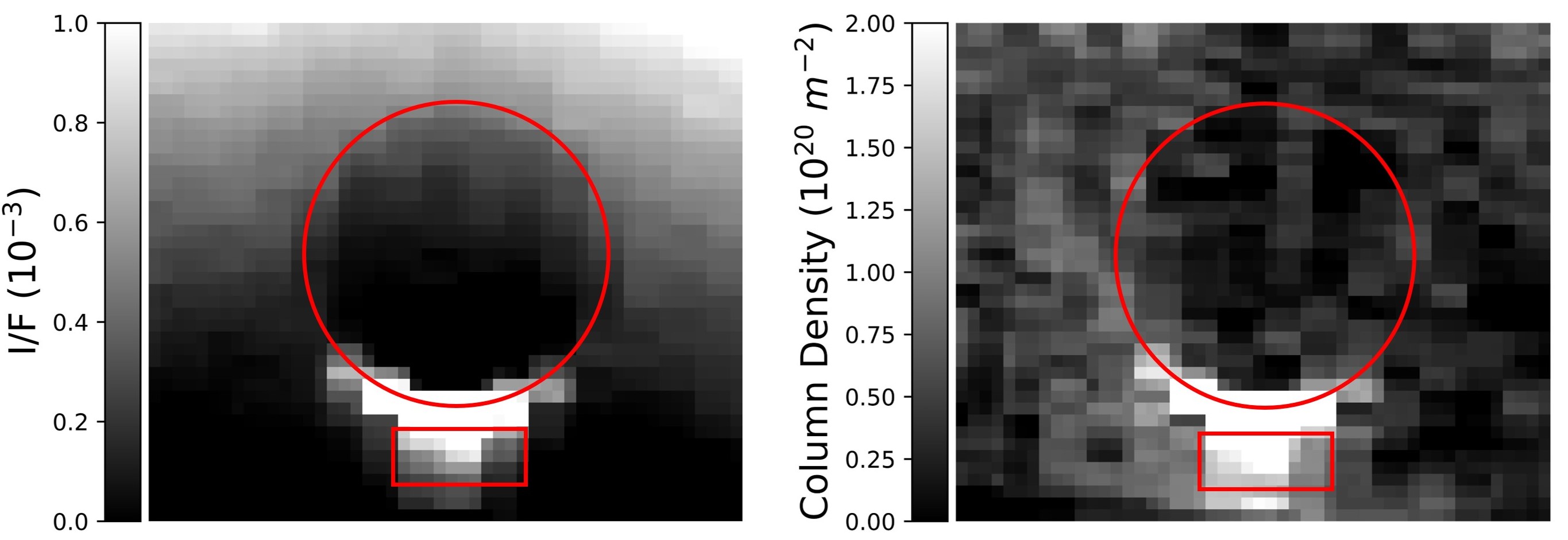}
\centering
\caption{\textit{Left:} Map of the plume's brightness at 2.70 $\mu$m derived from an average of five VIMS cubes (V1511798376, V1511800181, V1511800741, V1511801493, and  V1511802247). Note that this brightness is mostly due to reflected sunlight from Enceladus' lit limb and the plume particles. The image is re-projected on a grid with ~19 kilometers per pixel and oriented so that the plume points downward. \textit{Right:} Map of the water-vapor column density derived from the same averaged image. Column densities were calculated using the average residual values and Equation 1. Note that regions of high column density are visible outside the region where the plume particles are bright, which provides evidence that this signal traces a different plume component. The red circles outline Enceladus and the red rectangles show the pixels used for the spectral analysis (see Section~\ref{sec:data}).
\label{fig:Particle brightness & <n>}}
\end{figure*}

Enceladus' water-vapor flux can potentially be monitored over a broader range of conditions using near-infrared emission lines. Such lines were recently observed by the James Webb Space Telescope (JWST) \citep{Villanueva_23}, but there is evidence that VIMS can see them as well \citep{dhingra_17}. This paper will use the VIMS observations to constrain the water-vapor content of the plume over the course of the Cassini Mission. Section~\ref{sec:data} will discuss the format of the VIMS plume data and the methods we used to isolate the water-vapor signal. Section~\ref{sec:evidence} will demonstrate that the signal in the VIMS data is consistent with a real water-vapor feature using model-based predictions. Finally, we use this feature to calculate line-of-sight column density measurements over the course of the Cassini mission in Section~\ref{sec:varies}.

\section{Cassini VIMS Plume Data} \label{sec:data}
VIMS obtained observations of Enceladus and its plume at 352 wavelengths between 0.35 and 5.2 $\mu$m that are packaged as spectral cubes containing two spatial dimensions and one spectral dimension \citep{2004SSRv..115..111B}. We used calibration file RC19 and standard routines to convert the raw signal levels to reflectance in units of $I/F$. We computed geometrical parameters like Enceladus' orbital phase (i.e. mean anomaly) based on the timestamp for each observation. It is worth noting that a spectral shift of $\sim$10.4 $nm$ or 0.71 channels occurs over the course of the 13-year Cassini mission \citep{Clark}. This spectral shift is almost negligible for this study because the low signal-to-noise ratio and limited spectral resolution of VIMS cubes means that the narrow emission lines are not clearly resolved.

In order to facilitate comparisons among the various plume observations, the data from the infrared channel for each of these cubes were re-projected onto a uniform coordinate system aligned with Enceladus' spin axis and centered on the moon itself \citep{2013Natur.500..182H}. An example of the data found in these cubes is shown in Figure~\ref{fig:Particle brightness & <n>}. The \textit{Left} panel shows a map of the plume's brightness in the 2.60-2.75 $\mu$m wavelength band which encompasses the $\nu_{1}$ and $\nu_{3}$ vibrational bands of H$_{2}$O \citep{2015A&A...583A...6B}. This map is based on an average of five re-projected VIMS cubes and the bright particle plume is visible extending below Enceladus' south pole.  Note that these particular cubes were originally obtained in a mode where each pixel was 0.25 x 0.5 mrad \citep{2009ApJ...693.1749H}, so these pixels appear elongated in the horizontal direction after being re-projected.

While there are a few situations where the water-vapor signal is strong enough to be spatially mapped (Figure~\ref{fig:Particle brightness & <n>}: \textit{Right}, see Section~\ref{subsec:colden} for more details), for most of the VIMS observations we can only achieve sufficient signal-to-noise to quantify the water-vapor content of the plume by averaging all the data with the region containing the plume. To ensure consistency, we define the plume area as a region containing 55 pixels in the re-projected cubes covering altitudes $\sim$30 km to $\sim$130 km below Enceladus' South Pole and within $\sim$120 km of the moon's spin axis (see red rectangles in Figure~\ref{fig:Particle brightness & <n>}). These ranges were selected to avoid contamination from Enceladus' lit limb while still capturing altitudes where the water-vapor signal is expected to be strongest. In particular, the UVIS occultation data indicates that water-vapor column density is strongest at altitudes below 100 km \citep{2020Icar..34413461H}. To remove instrumental backgrounds, we subtract the mean signal in a nine pixel region centered at $\sim$330 km from Enceladus' spin axis and $\sim$20 km below the south pole. To filter outliers in each cube's average plume spectra, we also calculated standard deviations within each spectral channel and discarded $I/F$ values more than 2$\sigma$ away from the median when computing these averages. Finally, we exclude all data in spectral channel 2.564 $\mu$m because it consistently contains outliers throughout the data set even though this channel is not defined as a hot pixel in the VIMS documentation \citep{Clark}.

To ensure consistency among our estimates of the plume signal, we only considered cubes that fully covered the selected region and were obtained at distances within 300,000 km of Enceladus. There were 249 cubes spanning a range of orbital phases and observation times meeting these criteria, which are listed as groups in Table~\ref{tab:ColDenTab}. A list of the individual 249 spectral cubes can be found in the supplemental material. 

\section{Evidence of a Water Vapor Emission Feature} \label{sec:evidence}
The water-vapor signal in the VIMS data is relatively subtle because the instrument's limited spectral sampling ($\approx$ 16.6 nm/band) does not allow individual lines to be clearly resolved. Instead the spectral radiance appears as a slightly elevated brightness between 2.60 and 2.75 $\mu$m that is superimposed on top of a background spectral slope due to reflected sunlight from the particles in the plume and E ring \citep{dhingra_17}. We first noticed this slightly elevated brightness in these data in residuals from a Mie model fit to the spectra described in Appendix A. We therefore developed techniques to isolate the potential water-vapor signal from each cube that are described in Section~\ref{sec:models} below. In addition, we verified that the relevant spectral feature could indeed be due to water-vapor emission by both comparing the average signal to predictions from the Planetary Spectrum Generator (see Section~\ref{subsec:PSG}) and checking that the band area of this feature is consistent with the expected average column density of water vapor in the plume (see Section~\ref{subsec:colden}). 

\subsection{Isolating the water-vapor signal from the VIMS plume spectra} \label{sec:models}
\begin{figure}[h]
\includegraphics[width=0.5\textwidth]{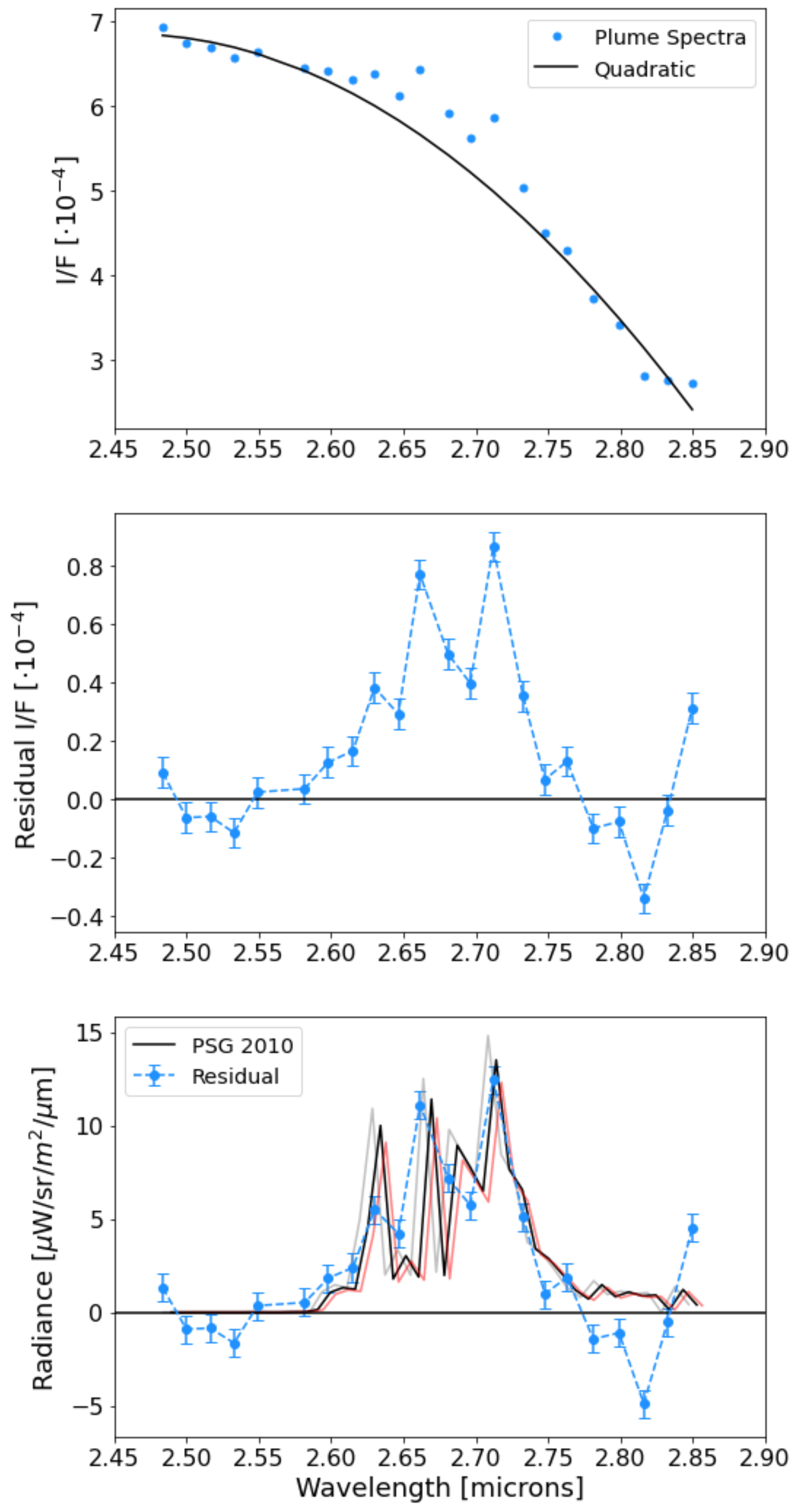}
\centering
\caption{The average plume spectra and residual of 249 spectral cubes from Cassini-VIMS observed from 2005 to 2015. The top panel shows the background-subtracted average spectra along with a quadratic fit to the regions between 2.484-2.598 $\mu$m and 2.748-2.850 $\mu$m. The middle panel shows the residual brightness after subtracting the quadratic polynomial. The bottom panel shows the same plume residual compared to the PSG models for 2005 (grey), 2010 (black), and 2015 (red). The wavelength shift in the PSG models was applied using values from \citet{Clark}. The models assume a column density of $3.63 \times 10^{19} m^{-2}$, consistent with the best-fit column density based on the feature's band area (see Section~\ref{subsec:colden}).
\label{fig:avg spectra plots}}
\end{figure}

For this analysis we will focus on the plume signals in the 2.484 $\mu$m - 2.850 $\mu$m wavelength range. These spectra include both the water-vapor signal between 2.60 and 2.75$\mu$m and background trends due to the plume and E-ring particles. These background trends can conveniently be approximated by a low order polynomial, so we can isolate the water-vapor signal by first defining three wavelength bands: 2.484-2.598$\mu$m, 2.615-2.733$\mu$m, and 2.748-2.850$\mu$m. The middle band contains the eight wavelength channels where we would expect to find the water vapor emission signal. In order to isolate this signal, we fit quadratic polynomials to the data in both surrounding bands (2.484-2.598 $\mu$m and 2.748-2.850 $\mu$m). These polynomial models are then subtracted from the spectra to produce residuals showing the intensity of water vapor emission in the middle band for each cube. (We also performed cubic fits, but found that these did not improve the results significantly.)
 
Figure~\ref{fig:avg spectra plots} (top panel) shows the non-weighted average spectra of the plume as derived from the entire 249 cube data set (Table~\ref{tab:ColDenTab}) and the resulting residual $I/F$ brightness (middle panel). Note that a straight average was taken because potential variations in the strength of the water-vapor feature (see Section~\ref{sec:varies}) are not properly accounted for using weights based on the error bars of individual spectral cubes. The error bars for this spectrum are therefore based on the root mean square residuals about the quadratic trend. The top panel of Figure~\ref{fig:avg spectra plots} shows a strong slope that is well fit by a quadratic model and can be attributed to the water-ice absorption band in the reflected sunlight from the particles in the plume \citep{2009ApJ...693.1749H,dhingra_17}. On top of this trend, a potential water-vapor signal is visible as a slight peak present around 2.60-2.75 $\mu$m. The middle panel shows the residual $I/F$ after subtracting the quadratic fit from the spectra, which isolates the potential water vapor emission feature. 

\subsection{Comparing the observed signal with predictions from the PSG} \label{subsec:PSG}
In order to determine if the residual signal shown in Figure~\ref{fig:avg spectra plots} is compatible with water vapor emission, we compared our observed spectra with predictions from the Planetary Spectrum Generator\footnote{https://psg.gsfc.nasa.gov} (PSG). This tool can model spectral features based on various properties related to the celestial body, detector, composition, etc \citep{2018JQSRT.217...86V}.  
For this particular study, we assume the Enceladus plume acts similarly to an expanding coma from a comet and consists of only water molecules since estimates of the water vapor abundance are within the 96-99$\%$ range \citep{postberg_2018}. We use a water-vapor column density of $3.63 \times 10^{19} m^{-2}$, which is consistent with calculations provided in Section~\ref{subsec:colden}. To estimate the temperature of the water vapor, we assume a rapid cooling off process occurs after the particles leave the tiger stripes where the interior temperature is estimated to be $197 \pm 20$K \citep{Goguen_2013}. Therefore, we use a temperature of 70K for the models because this is the equilibrium temperature of ice at Saturn's distance from the sun and it is consistent with the coldest temperatures seen on Enceladus' surface \citep{spencer_2018}. A full list of PSG input parameters can be found in Table~\ref{tab:PSG} in Appendix B. The lower panel in Figure~\ref{fig:avg spectra plots} shows PSG models from 2005, 2010, and 2015 compared to the average spectrum residual from all 249 spectral cubes (converted to units of Radiance). The PSG model has been applied three times to take into account the reduction of solar flux due to the increase of the heliocentric distance from 9.1 AU to 10 AU between 2005 and 2015. The spectral shift between the three PSG models shown in the plot corresponds to the wavelength shift in VIMS observations over the Cassini mission \citep{Clark}. A more detailed plot of this spectral channel shift can be found in Figure~\ref{fig:psg_shift} in Appendix B. Comparing the average residual signal to the three PSG models shows that the shape and intensity of the detected signal are consistent with predicted signal from water-vapor emission lines. This supports the hypothesis that this feature is a real signal from water vapor in the plume.

\begin{deluxetable*}{cccccccccc}[ht]
\tablenum{1}
\tablecaption{Average Water Vapor Column Density for Each Observation Group \label{tab:ColDenTab}} 
\tablehead{
\colhead{File} &  \nocolhead{} & \nocolhead{Cubes} & \colhead{Exposure} & \colhead{Observation} & \colhead{$\alpha$} & \colhead{Range} & \colhead{Orbital} & \colhead{$\langle$N$\rangle$\tablenotemark{e}} & \nocolhead{} \\
\colhead{Range} &  \colhead{Date} & \colhead{n\tablenotemark{a}} & \colhead{Time\tablenotemark{b}}& \colhead{Duration\tablenotemark{c}} & \colhead{(\degree)} & \colhead{(km)} & \colhead{Phase\tablenotemark{d}} & \colhead{($10^{19}$ m$^{-2}$)}  &  \colhead{$\chi_{red}^{2}$} \\
\nocolhead{} &\nocolhead{} & \nocolhead{} & \colhead{(Sec)} & \colhead{(Hrs)} & \nocolhead{} 
& \nocolhead{} & \colhead{(\degree)} & \nocolhead{} & \nocolhead{} 
}
\startdata
V1511794087-V1511810387	&	2005-11-27 &	14	& 7.9 &  4.6 &	161.2	&	141372.6	&	102.5, 151.0	&	9.34 ± 1.05	&	7.6	\\
V1556120297 	&	2007-04-24	&	1	& 0.4 & 0.1 &	155.4	&	184554.5	&	77.2	&	6.21 ± 4.40	&	\nodata	\\
V1569822293-V1569824909	&	2007-09-30	&	8	& 5.1 & 0.8 &	156.1	&	185659.5	&	288.8, 296.4	&	6.77 ± 1.19	&	4.4	\\
V1635804522-V1635817738	& 2009-11-01	&	22	& 14.1 & 3.7&	161.2	&	197151.8	&	153.0, 192.7	&	1.74 ± 1.14	&	10.9	\\
V1652826740-V1652830668	&	2010-05-17	&	12	& 7.7 & 1.1&	156.6	&	264376.5	&	7.6, 19.6	&	5.12 ± 0.97	&	4.8	\\
V1660403219-V1660407931	&	2010-08-13	&	9	& 5.8 & 1.4&	159.6	&	257005.5	&	336.5, 350.2	&	5.67 ± 2.58	&	3.2	\\
V1663878677-V1663881820	&	2010-09-22	&	12	& 7.7 & 0.9&	162.3	&	249462.4	&	92.2, 101.7	&	-2.95 ± 2.49	&	8.4	\\
V1671552668-V1671567368	&	2010-12-20	&	12	& 7.4 & 4.4&	157.6	&	133235.6	&	-1.7, 43.3 	&	0.29 ± 1.71	&	3.8	\\
V1675101079-V1675111889	&	2011-01-30	&	41	& 26.2 & 3.1&	152.6	&	215890	&	-27.3, 5.9	&	2.97 ± 1.07	&	7.2	\\
V1696147993-V1696155274	&	2011-10-01	&	24	& 15.4 & 2.1&	150.2	&	187076.3	&	171.3, 193.4	&	4.40 ± 0.75	&	9.3	\\
V1697685856-V1697694716	&	2011-10-19	&	19	& 12.2 &  2.5 &	150.2	&	189015.8	&	162.4, 189.1	&	4.90 ± 1.47	&	8.8	\\
V1699226646-V1699231874	&	2011-11-05	&	15	&  9.6 & 1.5&	149.3	&	190981.2	&	161.8, 177.5	&	6.29 ± 1.32	&	5.1	\\
V1711539317-V1711552021	&	2012-03-27	&	34	& 21.8 & 3.6 &	158.4	&	209498.1	&	115.2, 153.6	&	2.33 ± 0.85	&	7.4	\\
V1713081678-V1713090058	&	2012-04-14	&	21	& 13.4 & 2.4 &	158.2	&	208633.8	&	119.3, 144.6	&	3.14 ± 1.38	&	7.7	\\
V1823539171-V1823540036	&	2015-10-14	&	4	& 2.6  & 0.3 &	150.5	&	220495.3	&	54.2, 56.9	&	3.23 ± 4.45	&	11.2	\\
V1824760877 	&	2015-10-28	&	1	& 0.6 &  0.1 &	141.8	&	210623.6	&	164.4	&	21.0 ± 7.98	&	\nodata	\\
\enddata
\tablenotetext{a}{The number of spectral cubes in each observation group.}
\tablenotetext{b}{The sum of the total exposure time per pixel for n cubes in the sequence.}
\tablenotetext{c}{The duration window from the first observation to the last observation.}
\tablenotetext{d}{The minimum and maximum orbital phases for each observation group.}
\tablenotetext{e}{The weighted average group column density is 4.36 ± 0.32 $\times$ $10^{19}$ m$^{-2}$.}
\end{deluxetable*}

\subsection{Comparing water-vapor column densities with other measurements} \label{subsec:colden}
Another way to validate the water-vapor signal is to estimate the column density of H$_{2}$O molecules along the line of sight using the following equation \citep{2015A&A...583A...6B}.

\begin{equation}
\langle N \rangle = \frac{4\pi I_b}{h\nu g_f}
\end{equation}

where $I_{b}$ is the integrated band intensity $[W m^{-2} sr^{-1}]$ measured over the 2.615-2.733 $\mu$m wavelength range, $h$ is Planck's constant, $\nu$ is the central frequency of the band, and $g_{f}$ is the so-called g-factor or the band emission rate. For this particular band we used the g-factor from \citet{2015A&A...583A...6B}, which gives $g_{f}$ = 3.682 x $10^{-4}$ $s^{-1}$ at 1 AU. To correct for distance, we divided this g-factor by $r^{2}$ where $r$ is the range between Enceladus and the sun at the time of each observation in Astronomical Units. To convert our residual values to band area, we used the calibration equations from \citet{Clark} and solar flux values from the National Renewable Energy Laboratory \footnote{https://www.nrel.gov/grid/solar-resource/spectra-astm-e490.html} which are within 6.5$\%$ of Thekaekara's solar spectrum \citep{drummond_1973} used in the VIMS calibration pipeline. Solar flux values were corrected for the sun's range during each observation. 

The errors in the column density measurements are calculated using the standard deviation of residual spectral values in the surrounding wavelength bands (2.484-2.598$\mu$m and 2.748-2.850$\mu$m.) and dividing by the square root of the number of spectral channels in the middle band. This should provide a relatively conservative estimate of the uncertainty in the column density that is robust against uncertainties in the models for the background and the signal.

The right panel of Figure~\ref{fig:Particle brightness & <n>} shows the computed column density for the spectral image produced by averaging five data cubes (V1511800181 - V1511803001). This map shows that the water-vapor signal is more broadly distributed than the particle plume signal, being visible further below Enceladus and being noticeably wider than the particle plume signal visible in the left panel of the same figure. This distinct spatial distribution provides further evidence that this is a real water-vapor signal. 

In addition, the average spectrum shown in the bottom panel of Figure~\ref{fig:avg spectra plots} yields an estimated water vapor column density of $3.63 \pm 0.40 \times 10^{19} m^{-2}$, and for the individual observations the observed signals require column densities of order $10^{20}$ molecules/$m^2$ (see Section~\ref{sec:varies}). These values are consistent with column density measurements made from occultation observations from Cassini's Ultraviolet Imaging Spectrograph (UVIS) which gives values ranging from 9-15 $\times$ $10^{19} m^{-2}$ \citep{2020Icar..34413461H}. \citet{Villanueva_23} quotes an average column density of $1.7 \pm 0.1 \times 10^{18} m^{-2}$ from recent JWST observations of Enceladus. The JWST column density is about an order of magnitude lower than ours because the comparatively low spatial resolution of JWST (each pixel being several hundred kilometers wide) means that it can only measure average column densities over broad regions of the plume. By contrast, our data comes within a 100 km of the south polar terrain, where occultation measurements show the plume's column density is much larger \citep{2020Icar..34413461H}\footnote{Note that we do not provide estimates of out-gassing rates of the plume material because these calculations require secure measurements of how the vapor plume's brightness varies with altitude, which is hard to extract from our data due to limited signal to noise.}

\section{Water Vapor Variations} \label{sec:varies}
Having established that VIMS was able to detect the water vapor content of the plume, we can now use the VIMS observations to search for variations in Enceladus' water vapor activity on a range of different time scales. Residual spectra similar to Figure~\ref{fig:avg spectra plots} are computed on individual spectral cubes and used to estimate the number of H$_{2}$O molecules needed for each cube using the same procedures described in Section~\ref{subsec:colden}. As mentioned in Section~\ref{subsec:colden}, column density errors are calculated using the residuals from the quadratic fits to the background trend. We first use these estimates to examine whether the water vapor column densities vary systematically with orbital phase. Then we examine whether there could be variations in the column density on other time scales. 

\begin{figure*}[p]
\includegraphics[width=\textwidth]{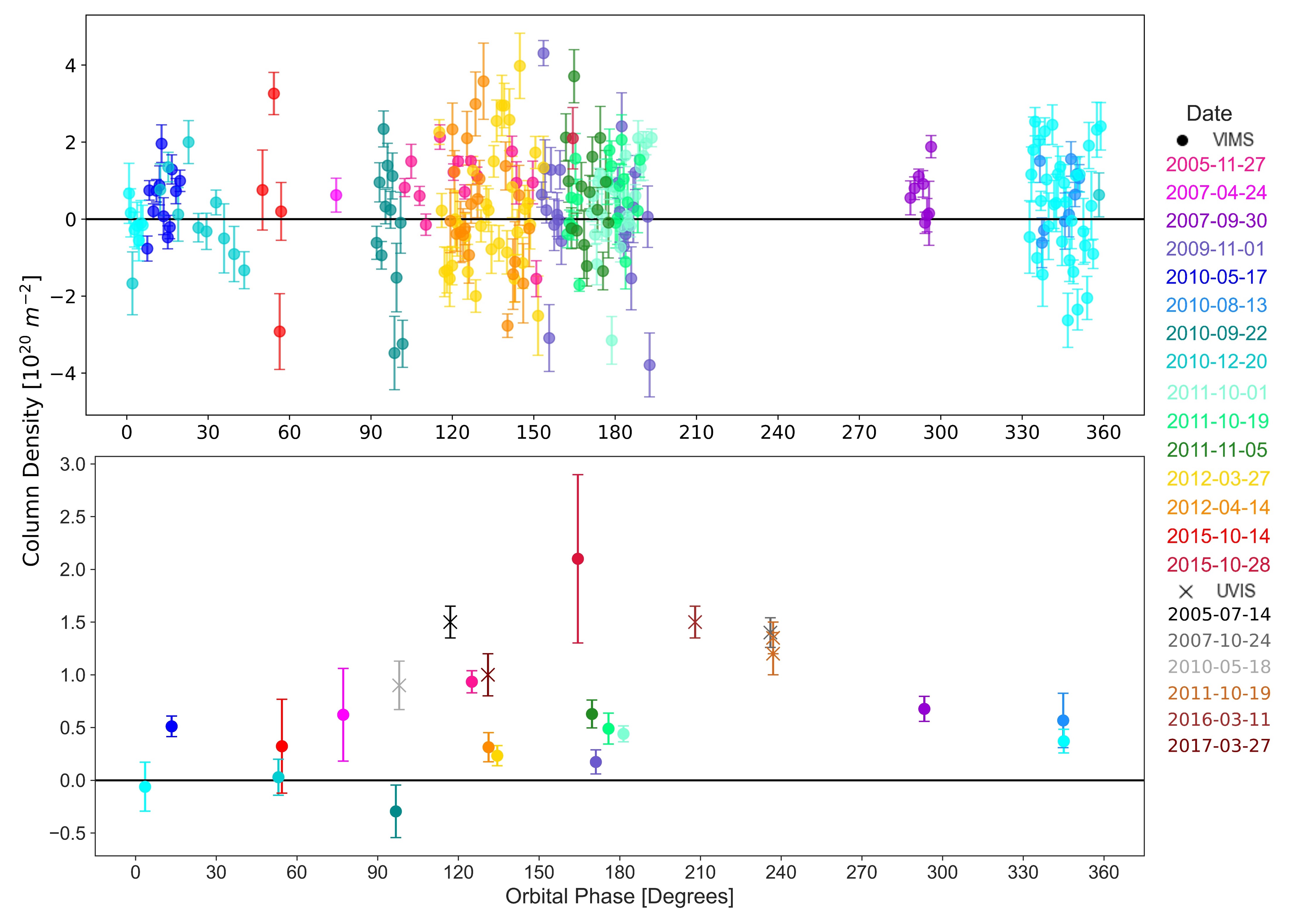}
\caption{Changes in water vapor column density as a function of Enceladus' orbital phase. The top panel shows the column density measurements for 249 spectral cubes with different colors representing the days in which the observations were taken. The bottom panel shows non-weighted average column density estimates for each observation group. Column density measurements of Enceladus plume occultations observed by UVIS obtained from \citet{2020Icar..34413461H} are also shown. There is no evidence for a clear pattern of water vapor column density variations over Enceladus' orbit. Note that although a small fraction of column densities in the top panel are negative, this is likely due to random noise because the signal-to-noise for individual measurements are relatively low. The fact that none of the averaged points in the lower panel are significantly negative is consistent with this idea.}\label{fig:variation plots}
\end{figure*}


\begin{figure*}[ht]
\includegraphics[width=\textwidth]{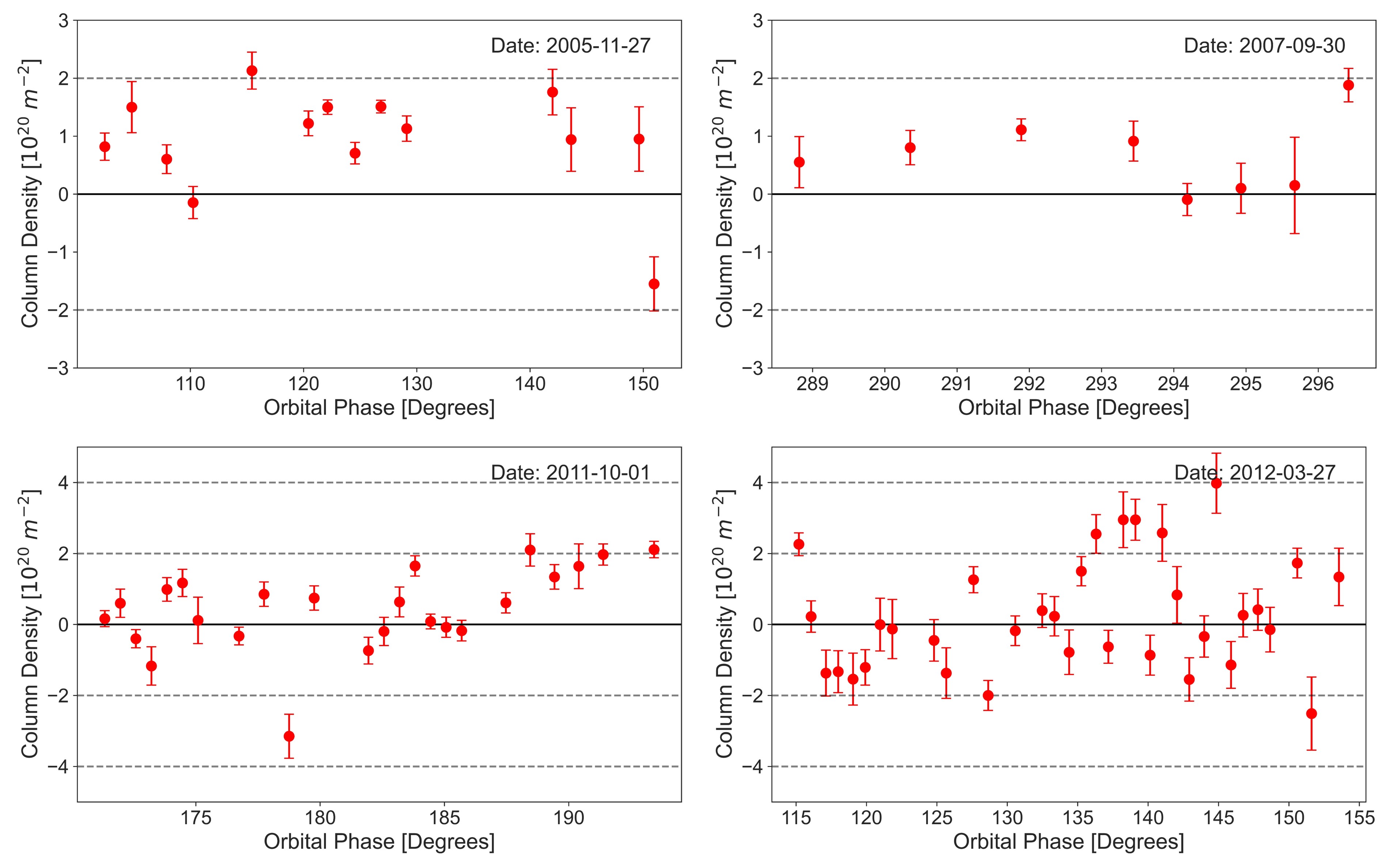}
\caption{Column density variations plotted over the orbital phase in four different observation groups. \textit{2005-11-27 Group:} Variations in the subset of observations analyzed in \citet{2009ApJ...693.1749H}. With good signal-to-noise ratios, this group consistently shows the water vapor signal in the middle five observations. \textit{2007-09-30 \& 2011-10-01 Groups:} Two examples of column density strengths possibly increasing or decreasing within one hour. \textit{2012-03-27 Group:} Represents an interesting pattern of column densities between orbital phases 135$\degree$ and 145$\degree$. This region appears to show the column density may be oscillating between two values, one around $3 \times 10^{20} m^{-2}$ and the other being within one sigma of 0. 
\label{fig: group plots}} 
\end{figure*}

Figure~\ref{fig:variation plots} shows the column density measurements for all 249 spectral cubes in our filtered data set as a function of orbital phase, and the observations are color-coded by the years they were observed. Note that while the data show substantial scatter, the column densities are on average positive, consistent with the detectable signal in the averaged data.  However, there is no clear pattern in column density fluctuations with orbital phase.

The lower panel of Figure~\ref{fig:variation plots} shows the variations of the non-weighted average column density estimates of groups of spectral cubes obtained within similar observation times (see average spectral residuals in Appendix D), along with the estimated water vapor column density measurements derived from UVIS occultation observations \citep{2020Icar..34413461H}. (Table~\ref{tab:ColDenTab} provides the average column density values for each of these observation groups, along with other properties of the data subsets.) The UVIS and VIMS estimates are generally consistent with each other (although the UVIS numbers may be a bit higher on average than the VIMS numbers) and neither shows clear evidence for systematic variations with orbital phase. If the water vapor flux varied with orbital phase like the particle flux, then the column density at orbital phases around 180$\degree$ (where Enceladus is furthest from Saturn) would always be about four times larger than it is at other orbital phases \citep{2020Icar..34413345I, 2014AJ....148...46N, Sharma_23}. We therefore agree with \citet{2020Icar..34413461H} that the water vapor plume does not show the same systematic activity variations as the particle plume.

While Figure~\ref{fig:variation plots} indicates that there are no obvious trends with orbital phase, it is also important to look for evidence of longer-term variations among the different observations and short-term variations within each observation group. The column density estimates shown in Table~\ref{tab:ColDenTab} and the average residual signals for individual groups shown in Appendix D may reveal hints of yearly variations over the Cassini mission. The water-vapor flux seems to have been particularly high in the VIMS observations taken in 2005 and 2007. More explicitly, there are roughly eight groups that have 1$\sigma$ error bars on the order of $1\times 10^{19}$ m$^{-2}$ (see Table~\ref{tab:ColDenTab}). Among these higher-quality observations, groups 2005-11-27 and 2007-09-30 have estimated column densities that are $\sim$5$\sigma$ above zero and show signals around 2.7 $\mu$m that are consistent with those expected for the water vapor flux feature (see Appendix D). By contrast, groups 2011-01-30, 2011-10-01, and 2012-03-27, despite also having error bars around $1\times 10^{19}$ m$^{-2}$, have lower column densities, with two of them being less than 3$\sigma$ above zero. Assuming our error bars are accurate, this decline is marginally statistically significant because the average signal of the 2005-11-27 and 2007-09-30 observations is roughly $\sim$3$\sigma$ above the average signal of the 2011-01-30, 2011-10-01, and 2012-03-27 observations.  This implies that we have either underestimated our error bars by roughly a factor two, or that the typical vapor flux may have been lower during the observations made in the latter part of the Cassini Mission.

In order to investigate shorter-term variations in the plume, we examined the column density estimates as a function of the orbital phase for each observation group. First, we calculated reduced $\chi^{2}$ values for each observation group's estimated column density, which are provided in Table~\ref{tab:ColDenTab}. Most of these reduced $\chi^{2}$ values greatly exceed 1, implying that the scatter among the observations within each group is roughly 2-3 times larger than expected given the conservatively-defined error bars. For all groups consisting of multiple spectral cubes, the probabilities to exceed associated with these reduced $\chi^{2}$ values is equal to or much less than 0.01$\%$. It is therefore worth taking a closer look at these sub-sets of the data to see if there are any trends that could represent short-term variations. Figure~\ref{fig: group plots} illustrates the water vapor variations for four particularly interesting observation groups in our data set (The remaining observation group plots can be found in Appendix C). 

First, the 2005-11-27 group plot shows observations that were previously studied in great detail in \citet{2009ApJ...693.1749H} due to their relatively high signal-to-noise ratios (note that four out of the five observations between orbital phases of 120$\degree$ and 130$\degree$ show the water vapor emission signal clearly in their individual spectra.) While the signal appears to be present in the majority of these cubes, the scatter among the observations is larger than their formal error bars, and the signal between 120$\degree$ and 130$\degree$ is on average higher than the observations between 100$\degree$ and 110$\degree$.

The 2007-09-30 group shows evidence for a possible dip in the water-vapor flux between 294$\degree$ and 296$\degree$, which spans a time of approximately 10 minutes. This feature appears unlikely to just be a statistical artifact because it is coherent and localized in time. Hence, unless this is due to some unidentified systematic artifact in the data, then it represents a very rapid fluctuation in the plume’s water vapor flux. 

The 2011-10-01 group shows a possible increase in the vapor plume activity between 185$\degree$ and 195$\degree$ in a duration of 30 minutes. If this increase is real, the rise in water-vapor flux after Enceladus passes apocenter is interesting because the particle flux would be decreasing in this orbital phase range \citep{2020Icar..34413345I,Sharma_23}.

The 2012-03-27 group maps an interesting result that occurs between orbital phases 135$\degree$ and 143$\degree$. The water-vapor output tends to bifurcate in this region between relatively large column density measurements and negative ones. Analysis of the individual spectra of the observations in this orbital phase range shows no clear cause for this odd pattern. While the signal-to-noise ratios in this observation group are fairly low, some spectra exhibit clear emission peaks while others do not.

These data therefore provide some evidence that the plume's vapor activity can potentially vary on short timescales. Unfortunately, most of our groups do not have high enough signal-to-noise ratios to adequately map such small scale variations (see Figure~\ref{fig: group plots}: 2012-03-27), and even the variations shown in Figure~\ref{fig: group plots} look to be somewhat stochastic. However, given we used relatively conservative error bars, our analysis suggests that $\leq$1 hour variations in water vapor column density cannot be excluded, and merit further investigation, especially if additional observations from JWST become available. 

Finally, we can note that that the total data set of column densities shows an excess variance similar to or smaller than the variance of the individual groups (the overall reduced $\chi^2$ for the entire data set being around 4). This result implies that while there are significant variations among the groups, these fluctuations are smaller than the variations within most of the observation groups, and so the variations among the groups could potentially be due to residual short-term variations. 

\section{Discussion}
This analysis provides strong evidence that the 2.7 micron water-vapor emission feature is detectable in the Cassini-VIMS spectra of Enceladus' plume. These data are consistent with previous UVIS measurements of the plume's vapor column density, and also confirm that the plume's vapor content does not vary dramatically with orbital phase. However, we also find hints that the plume's vapor content may vary on time scales as short as around an hour.

The results of this study confirm that there is a difference between the tidal variation of the plume's particle flux and the fluctuations in the plume's vapor content. This suggests that the variation in the water-vapor plume is complicated by other geological processes occurring within the fissures. Therefore, our detections of minimal variation over orbital phase and possible small-scale variations in observation groups could be consistent with individual plume jets turning on and off as suggested by \citet{2020Icar..34413461H} and studied by \citet{Portyankina_22}. This analysis could potentially provide constraints on the frequency and durations of these jet events. Now that JWST has demonstrated that it can detect Enceladus' vapor plume \citep{Villanueva_23}, future JWST observations could contribute a deeper understanding of the fluctuations of water-vapor content. 

\section*{Acknowledgments}
This research was supported by NASA's Cassini Data Analysis Program (CDAP) grants 80NSSC18K1701 and 80NSSC21K0427. We would like to thank Dr. Shannon MacKenzie, Dr. John Spencer, Dr. Sanaz Vahidinia, and Dr. Jay Goguen for collaborating and offering insights into this project. 

\pagebreak
\section*{Appendix A: Mie Models} \label{app:mie}
\begin{figure}[ht]
\includegraphics[width=0.46\textwidth]{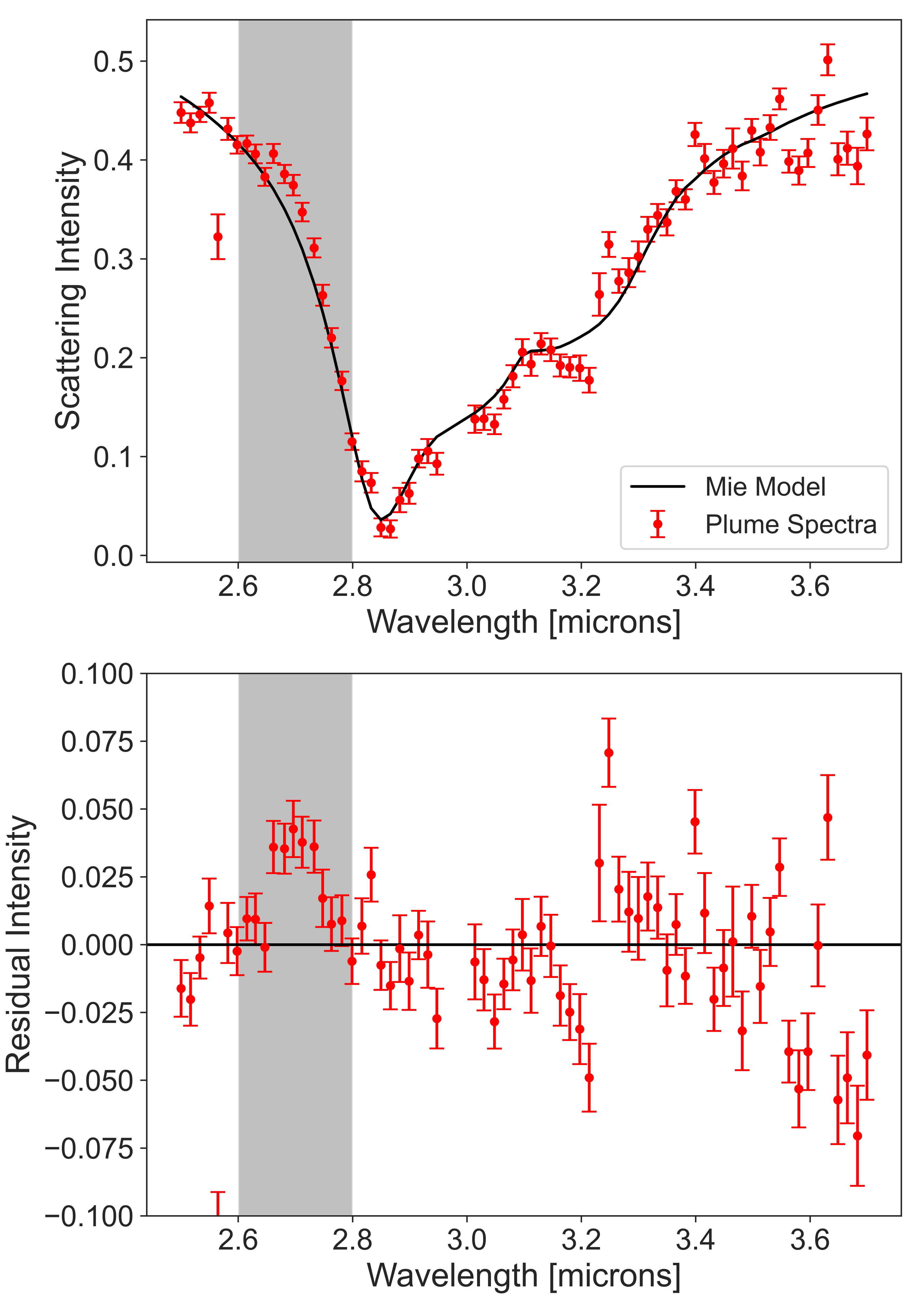}
\centering
\caption{Modeling Enceladus plume spectra using Mie Scattering theory. The processed plume spectra were obtained from \citet{2009ApJ...693.1749H}. \textit{Upper:} shows the best-fit Mie Model for plume spectra observed 35 km above the surface of Enceladus in the 2.5-3.7$\mu$m wavelength range. \textit{Lower:} shows the resulting residual intensity after subtracting the Mie model from the spectra. The shaded regions on each plot highlight the wavelength band of interest, a near-infrared region that contains a water vapor emission peak.
\label{fig:Mie plots}}
\end{figure}

The first evidence of the water-vapor feature came from examining residuals of Enceladus plume spectra from Mie Scattering models. Mie Scattering models are used for various applications including modeling coma particles from comets \citep{2020Natur.578...49F}. In this study, the Mie models were utilized to model the crystalline water-ice 3.0 $\mu$m absorption feature in the particulate component of the Enceladus plume. We modeled a data set obtained from \citet{2009ApJ...693.1749H} containing spectra of effective altitudes above the surface of Enceladus processed from a subset of 14 spectral cubes observed in 2005. 
 
In fitting these models, a slight peak appears in the 2.60-2.75 $\mu$m range in some of the spectra, while usually showing up in spectra observed at lower altitudes. The upper panel in Figure~\ref{fig:Mie plots} shows the best-fit Mie model for the plume spectra at an altitude of 35 km in the 2.50-3.70 $\mu$m range. The lower panel in Figure~\ref{fig:Mie plots} shows the residual brightness produced by subtracting the Mie model from the plume spectra which shows a clear peak. This finding prompted us to expand our analysis to a larger data set of VIMS spectral cubes and develop quadratic fitting models for isolating the water-vapor emission feature from the water-ice absorption spectra. 

\newpage
\section*{Appendix B: PSG} \label{app:psg}
\begin{figure*}[hbt!]
\centering
\includegraphics[width=0.8\textwidth]{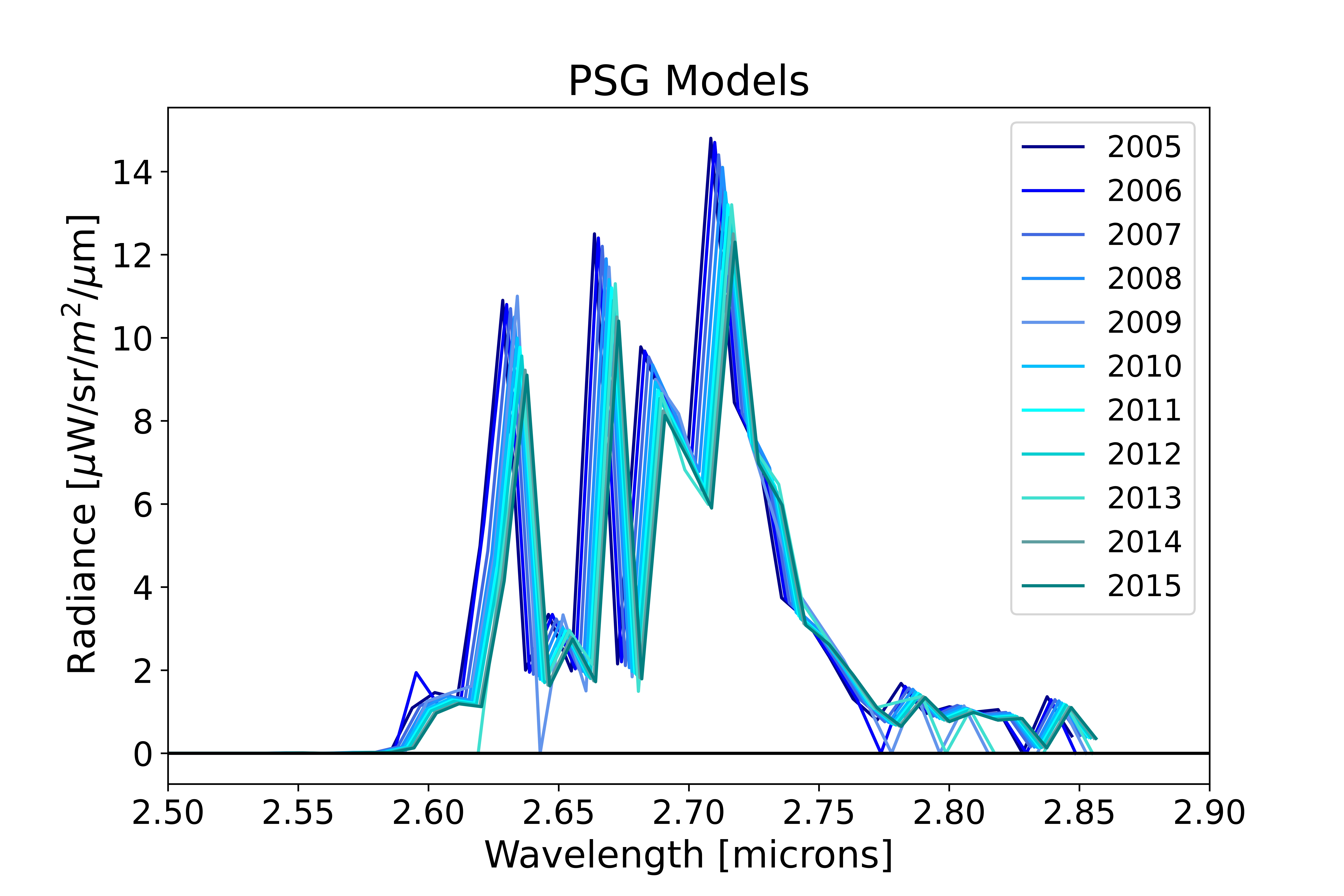}
\caption{Theoretical spectral channel shift of the water vapor emission feature in Cassini-VIMS observations from 2005 to 2015. Spectral shift values were taken from \citet{Clark}, and the spectral models were produced using the PSG.  
\label{fig:psg_shift}} 
\end{figure*}

Table~\ref{tab:PSG} shows the parameters used in this study to produce PSG models analyzed in Section~\ref{subsec:PSG}. Due to the structure of Enceladus' plume and the composition of the plume material, we modeled the atmosphere of Enceladus as an expanding coma. Figure~\ref{fig:psg_shift} shows the shift in spectral channels in Cassini-VIMS observations from 2005 to 2015. The shift is represented by ten PSG models of the water vapor emission feature between 2.60 and 2.75 microns. This shift is documented in detail in \citet{Clark}. The decrease in intensity over this time period is due to the increasing distance between the sun and the Saturnian system as Saturn completes its 29-year orbit. 

\begin{deluxetable}{lll}[h]
\tablenum{2}
\tablecaption{Planetary Spectrum Generator Parameters \label{tab:PSG}}
\tablewidth{0pt}
\tablehead{\colhead{Parameter} & \colhead{Value}
}
\startdata
Object	&	S-Enceladus	\\
Viewing Geometry	&	Limb	\\
Altitude	&	1000 $km$	\\
Atmosphere	&	Expanding Coma	\\
Activity\tablenotemark{a}	&	3.63 x $10^{19}$ m$^{-2}$	\\
Temperature\tablenotemark{b}	&	70 K	\\
Expansion Velocity\tablenotemark{c}	&	780 $m/s$	\\
Composition	&	GSFC[H2O]	\\
Abundance	&	100$\%$	\\
Spectral Range	&	2.5-2.9 $\mu$m	\\
Resolution (Resolving Power)\tablenotemark{d}	&	300	\\
Include Molecular Signatures	&	(Check)	\\
Spectrum Intensity Unit	&	W/sr/m$^{2}$/$\mu$m \\
\enddata
\tablenotetext{a}{Column density estimate from Section~\ref{sec:evidence}.}
\tablenotetext{b}{Estimated temperature of Enceladus' surface. This is different from the rotational temperature of water molecules as quoted in \citet{Villanueva_23}.}
\tablenotetext{c}{Expansion velocity value taken from \citet{Villanueva_23}.}
\tablenotetext{d}{Resolving power value taken from \citet{2004SSRv..115..111B}.}
\end{deluxetable}

\clearpage
\onecolumngrid
\section*{Appendix C: Column Density Plots} \label{app:colden}
Similar to Figure~\ref{fig: group plots}, the figures below show the column density variations of the rest of the observation groups excluding the groups with only one spectral cube. Other details of these groups can be found in Table~\ref{tab:ColDenTab}. 

\begin{figure*}[h]
\centering
\includegraphics[width=0.89\textwidth]{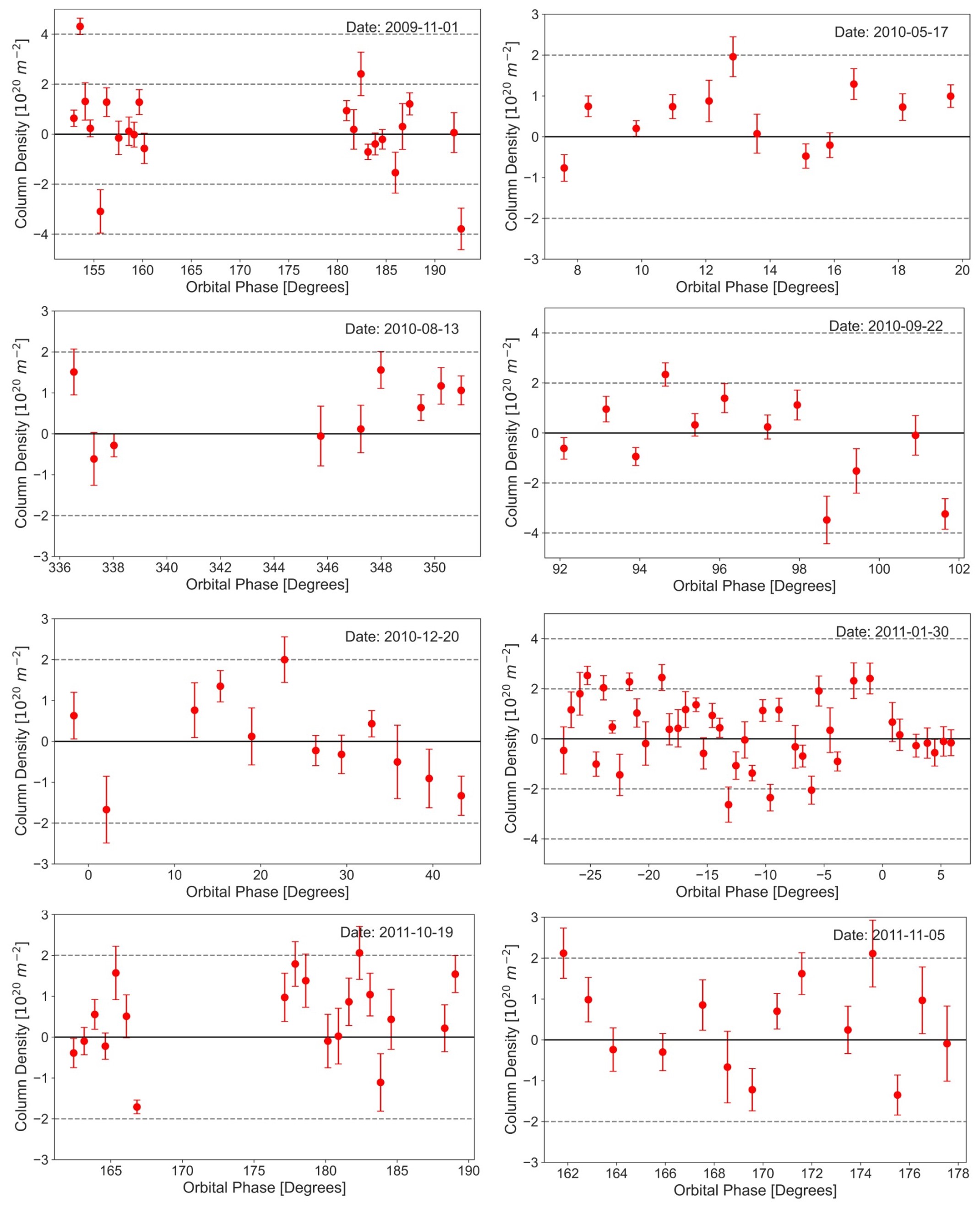}
\caption{Column density variations for groups 2009-11-01 through 2011-11-05. Variations in groups 2005-11-27, 2007-09-30, and  2011-10-01 can be found in Figure~\ref{fig: group plots}.
\label{fig: extra plots1}}
\end{figure*}

\begin{figure*}[ht!]
\centering
\includegraphics[width=0.89\textwidth]{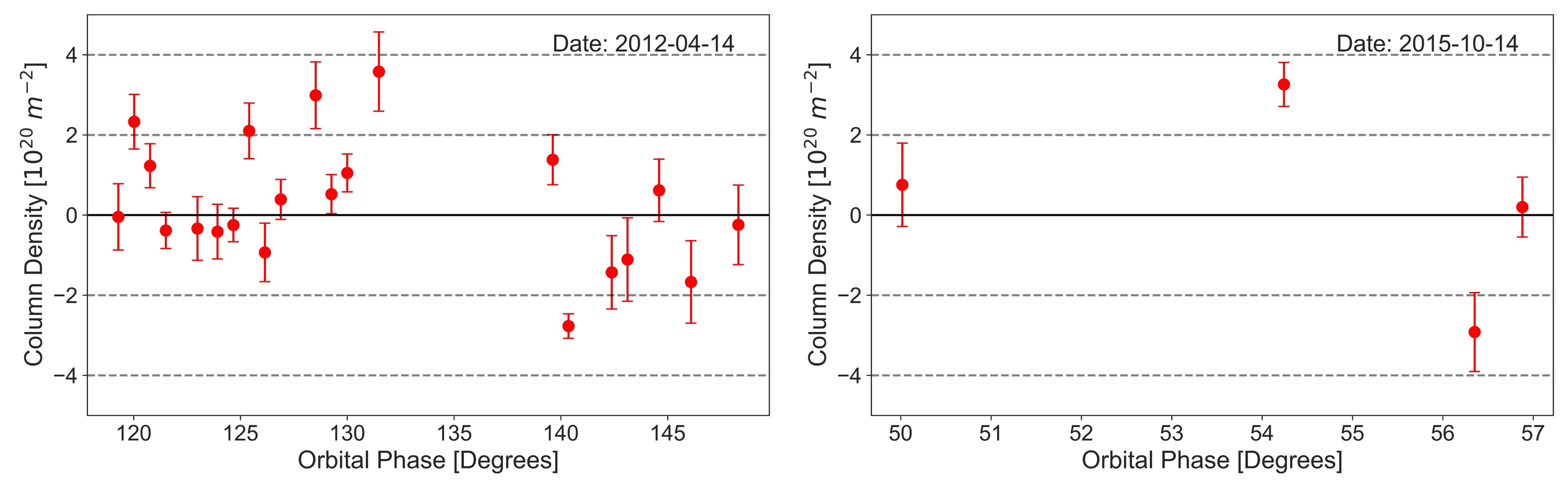}
\caption{Column density variations for groups 2012-04-14 and 2015-10-14. Variations in group 2012-03-27 can be found in Figure~\ref{fig: group plots}.
\label{fig: extra plots2}}
\end{figure*}

\section*{Appendix D: Residuals} \label{app:res}
\begin{figure*}[ht!]
\centering
\includegraphics[width=0.9\textwidth]{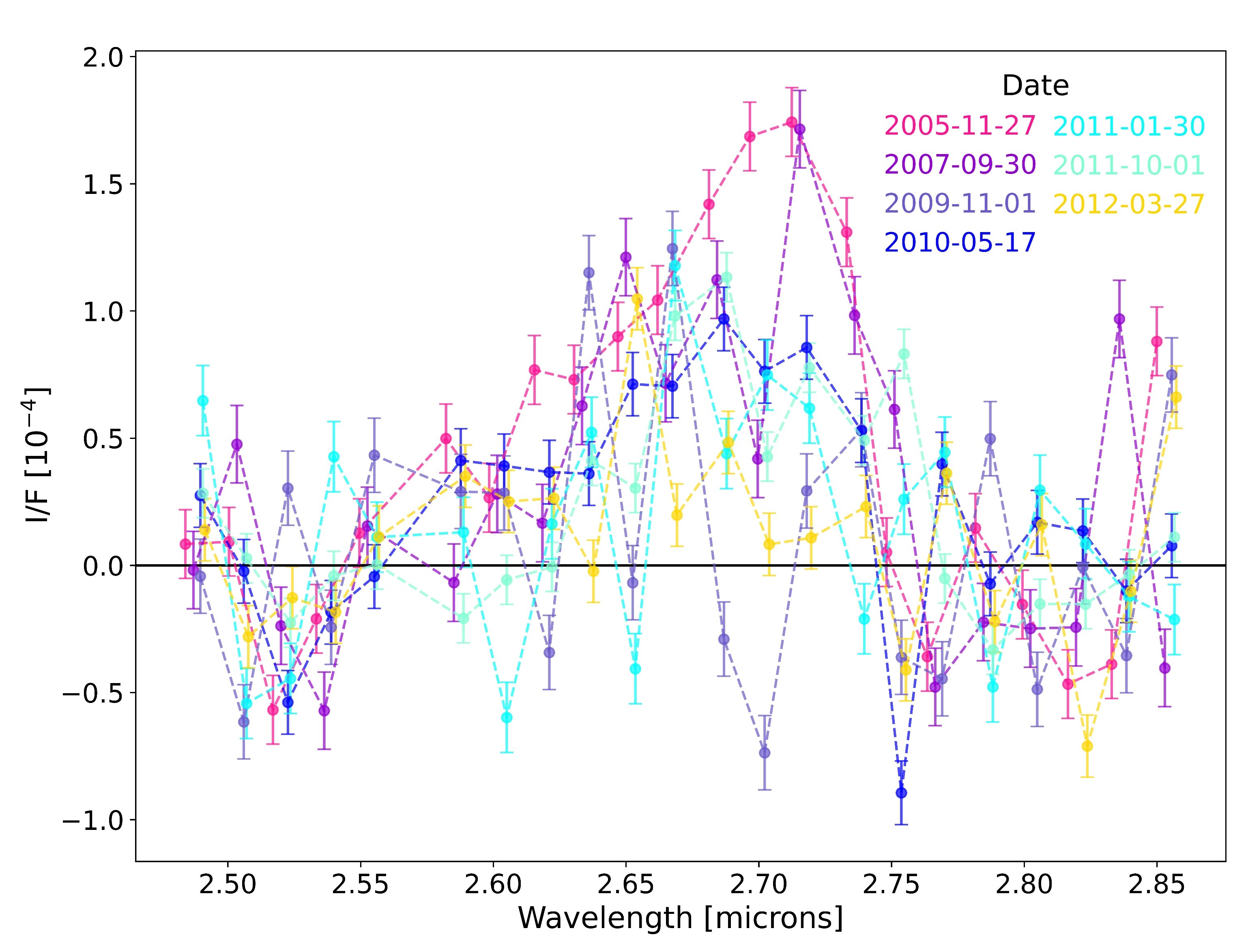}
\caption{The non-weighted average spectral residuals for observation groups with a column density error lower than 1.2$\times$ $10^{19}$ $m^{-2}$. Each residual is colored by the date the observations were taken corresponding to Figure~\ref{fig:variation plots}.
\label{fig: grouped_res}}
\end{figure*}

\begin{figure*}[p]
\centering
\includegraphics[width=0.85\textwidth]{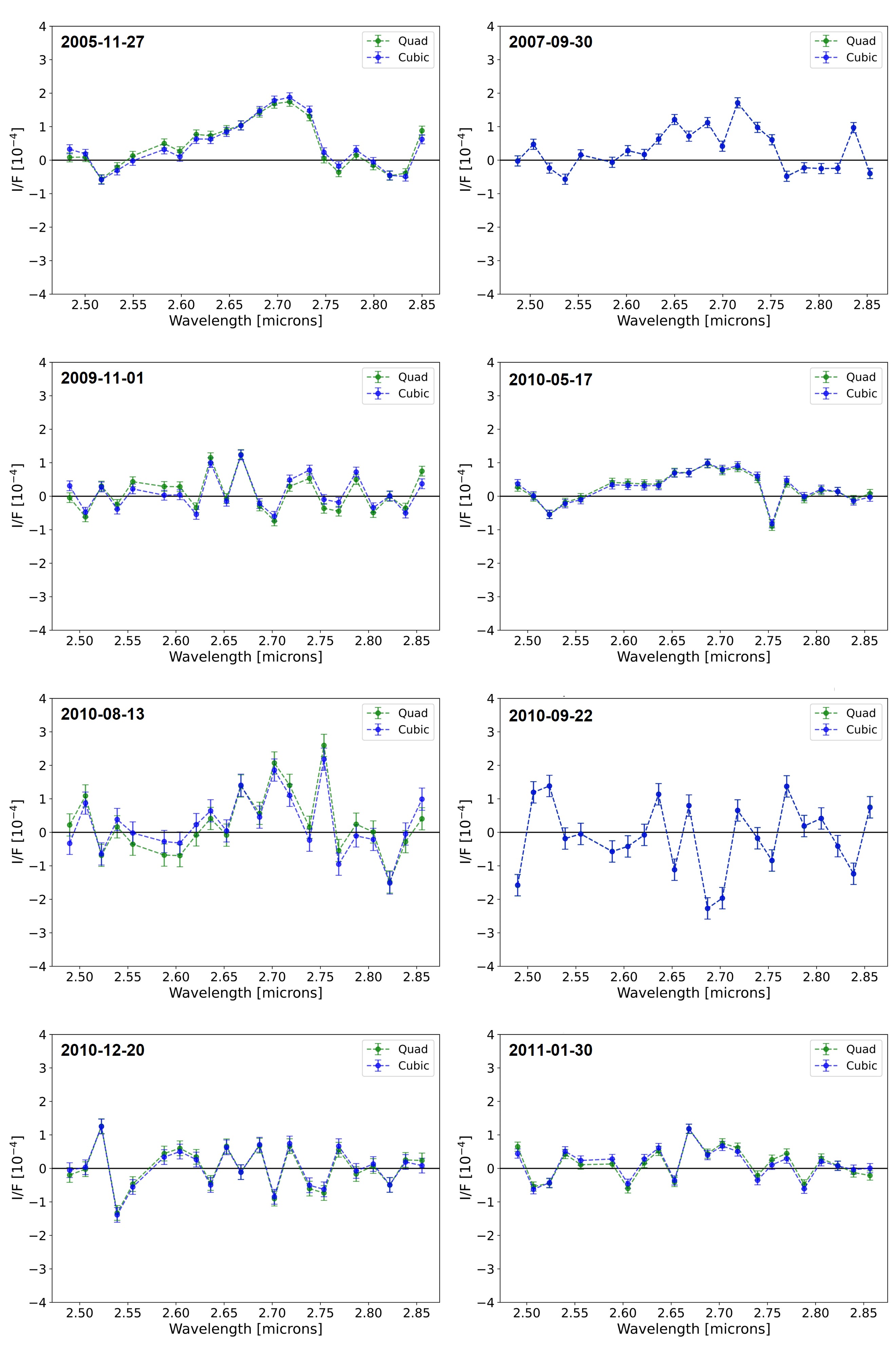}
\caption{Residual Spectra for groups 2005-11-27 through 2011-01-30.
\label{fig: res plots1}}
\end{figure*}

\begin{figure*}[ht!]
\centering
\includegraphics[width=0.85\textwidth]{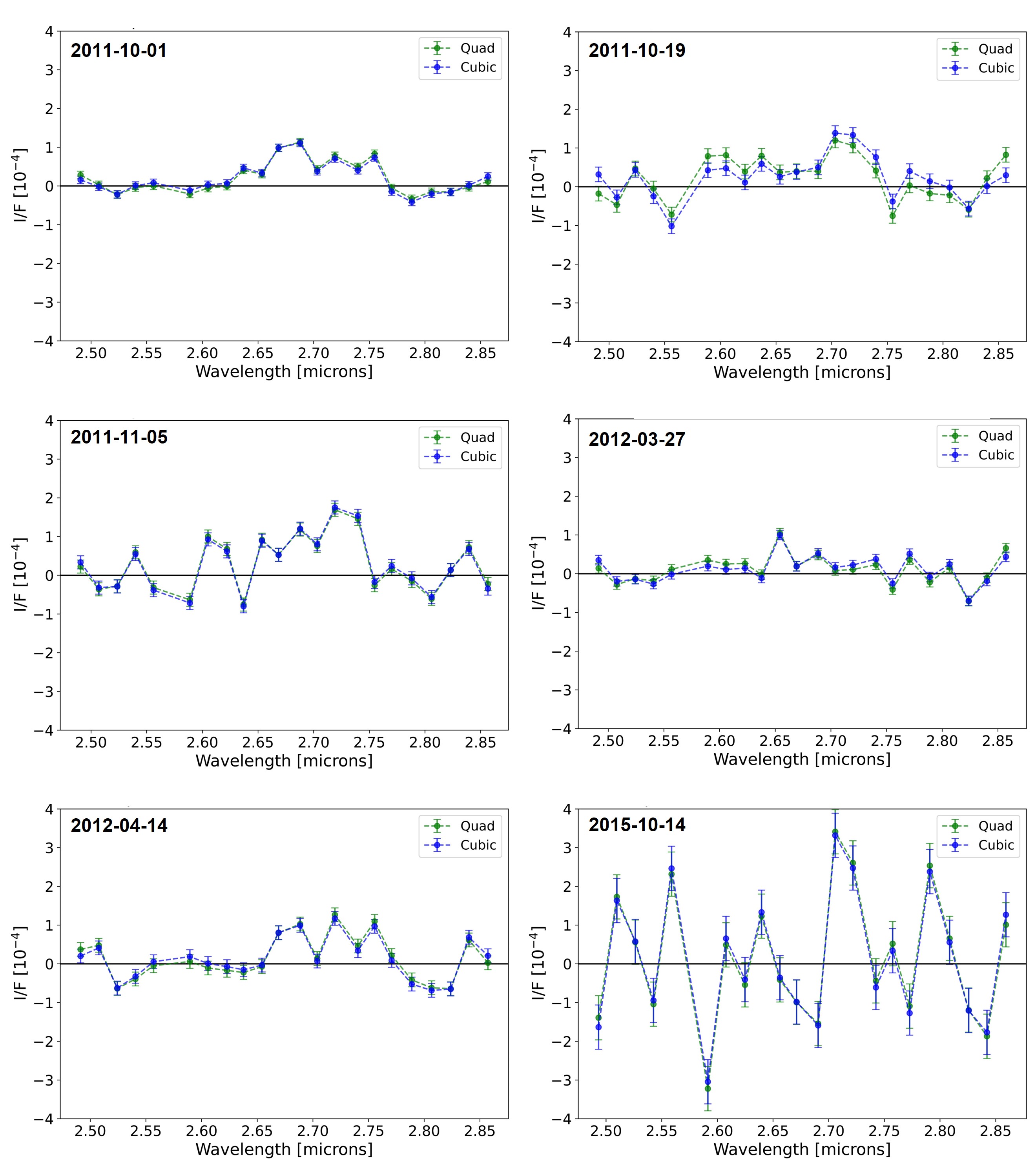}
\caption{Residual Spectra for groups 2011-10-01 through 2015-10-14.
\label{fig: res plots2}}
\end{figure*}

Figure~\ref{fig: grouped_res} shows multiple residuals obtained from averaged spectra of spectral cubes with similar observation times (see Table~\ref{tab:ColDenTab}). The spectral residuals shown are associated with column density uncertainties lower than 1.2 $\times$ $10^{19}$ $m^{-2}$. Figures~\ref{fig: res plots1} and \ref{fig: res plots2} show individual spectral residuals for each observation group listed in Table~\ref{tab:ColDenTab}.

\FloatBarrier
\clearpage
\bibliography{draft}{}
\bibliographystyle{aasjournal}

\end{document}